# Quantized Conductance and Switching in Percolating Nanoparticle Films

Abdul Sattar, Shawn Fostner and Simon A. Brown*

*The MacDiarmid Institute for Advanced Materials and Nanotechnology, Department of Physics and Astronomy, University of Canterbury, Private Bag 4800, Christchurch 8140, New Zealand.*

We demonstrate switching behavior and quantized conductance at room temperature in percolating films of nanoparticles. Our experiments and complementary simulations show that switching and quantization result from formation of atomic scale wires in gaps between particles. These effects occur only when tunnel gaps are present in the film, close to the percolation threshold.

*E-mail: simon.brown@canterbury.ac.nz

It is now 25 years since the discovery that the conductance of small wires can be quantized, [1,2] but interest in quantized conduction is still high because of both intriguing fundamental physical phenomena [3,4] and potential technological applications. In particular, the switching behaviour that occurs when a device steps between different quantized conductance levels could be used in new memory architectures [5], as well as in related neuromorphic [6], and memristor [7, 8] devices.

For useful applications it is essential that devices exhibit quantized conductance and switching at room temperature (rather than cryogenic temperatures). This is extremely challenging. The most successful approach is based on the use of solid electrolytes [5], but these devices rely on chemical processes which occur in only a small number of materials systems.

Here we show that percolating systems [9] of randomly deposited nanoparticles (Fig. 1 (a)) exhibit room temperature quantized conductance and switching, both during, and in response to applied voltages. Our complementary simulations clearly demonstrate that this behavior is due to formation of atomic scale wires within gaps in the film of particles. This easily fabricated and apparently random system can yield both stable long-lived structures, and results that are similar to those from single junction systems.

Quantisation of the conductance of a wire is a signature of quantum confinement i.e. when the width of a wire is sufficiently small (comparable to the Fermi wavelength ($\lambda_F$) of the carriers), quantized energy levels are formed, and the conductance of the wire is an integral multiple of $G_0 = 2 e^2/h$. [10] In semiconductor devices $\lambda_F$ can be relatively large, so that quantized conductance can be observed in devices with dimensions of hundreds of nanometers, at cryogenic temperatures. In metals $\lambda_F$ is typically so small that quantized conductance can only be observed in atomic scale structures. Mechanically controlled breakjunctions (MCBJs) [11] are perhaps the most popular method for achieving the required atomic scale structures, and allow quantized conduction to be observed even at room temperature, but typical lifetimes are only milliseconds.[11,12] The behaviour of MCBJs can sometimes be emulated via electromigration processes in metallic devices,[13] but in many cases quantization is not observed.[14]

We randomly deposit pre-formed 7nm tin clusters between pairs of electrical contacts. [15,16] The clusters overlap and coalesce [17,18] when they land on each other and hence form connected groups of particles that grow in lateral extent as more clusters are deposited. The coalescence ensures that the coupling between particles (or grains) is very strong and that (at least in the regime of interest, near the percolation threshold and near to room temperature) the groups behave as metallic objects with negligible charging energy and level spacing. [19] In the absence of connections between groups of particles tunnel gaps determine the properties of the devices (see below) and the devices are best understood within a framework of percolation and tunneling (Fig. 1 (a)) rather than one of granular films where Coulomb charging and the discrete single particle energy level spectrum are important. [19]

By monitoring the conductance of the devices during deposition[15] we are able to ensure that the final surface coverage $p$ is close to the percolation threshold $p_c$ ~ 0.68. [20] We focus on results from 67 samples with a morphology (Fig. 1(b)) which is optimal for observation of quantized conduction and switching behavior.

Fig. 1 (c) shows the onset of conduction for a typical sample. The conductance clearly exhibits steps that coincide with integral multiples of $G_0$. Fig. 1 (d) shows a histogram of conductance values measured during the onset of conduction, showing clear peaks at integral multiples of $G_0$. This histogram is strikingly similar to the histograms obtained from MCBJs. [11] Every one of the 67 samples with the optimal morphology (Fig. 1 (b)) exhibited conductance steps at the onset of conductance and we emphasize that the same quantized conductance values are observed independent of sample area, aspect ratio, film thickness and applied voltage.

It is well established in MCBJs that there are a number of factors (alignment with the contacts, disorder) that result in transmission coefficients [10,11] that are different from unity and that therefore lead to system conductances that differ from integral multiples of $G_0$. Hence quantized conductance is observed most clearly in histograms like Fig. 1 (d). The Fermi wavelength of tin is 0.337 nm ($E_F$ = 10.2 $eV$, $m^*$=1.3 $m_e$), [21] which is comparable to the atomic diameter, and so quantization of the conductance is clear evidence for the formation of atomic scale wires within our samples.

We now focus on the mechanism of formation of the atomic wire in the gap between particles. When two surfaces approach each other to within a few Å, attractive forces cause an instability [22] and can lead to a 'jump to contact' in which an atomic scale wire is formed. [11,23,24] During deposition the particle film initially contains large tunneling gaps but as the coverage approaches the percolation threshold the average size of the groups of particles grows, and the number and size of the gaps decreases. When the gap size becomes small enough a jump to contact, which creates an atomic scale wire, can occur. This picture is consistent with the deposition experiments, since the conductance steps stop if the deposition is interrupted, and start again if deposition is restarted.

Insight into the microscopic mechanism for the jump to contact is provided by Ref. [25] which shows that high electric fields are required for electric field induced evaporation (EFIE

– Fig. 1 (e)) and electric field induced surface diffusion (EFISD – Fig. 1 (f)), whereas (independent of applied electric field) attractive van der Waals forces can cause jump to contact at interparticle distances smaller than ≤1.5 nm. Since very low DC voltages (<50mV) are applied during deposition it is clear that the electric fields in even the smallest gaps are insufficient for EFIE and EFISD. [25,26,27] Hence, attractive van der Waals forces are responsible for wire formation during the onset of conduction.

We now turn our attention to switching behaviour and quantization in the same samples after deposition. The samples are quite stable as long as only a small constant DC voltage is applied, but Fig. 2 (a) shows significant stepwise changes in the conductance of a 10 μm x 20 μm sample when a larger voltage is applied: the conductance exhibits 3 distinct steps with values that are close to integral multiples of $G_0$.

Small (10μm x 20μm) samples exhibit only stepwise *decreases* in conductance (as in Fig. 2 (a)), whereas large (100μm x 200μm) samples exhibit a more complex series of conductance steps, with *both* increases and decreases of the conductance (Fig. 3 (a)). Repeated steps are often observed between similar conductance levels, while in other cases the conductance steps between multiple levels. This behavior is observed because in large samples there is typically a more complex network, with a much larger number of percolating-tunneling pathways (solid and dotted lines in Fig. 1 (a)) compared to the small samples.

A histogram of observed conductance values during voltage ramps for all 67 samples (Fig. 2 (b)) shows a clear peak at $G_0$, and much weaker peaks centered near $2G_0$ and $3G_0$. This quantization behavior is very similar to that observed in MCBJs. [11] It is occasionally observed (Fig. 3 (b)) that the size of the step (rather than its absolute conductance) is quantized; this is consistent with the formation of an atomic scale wire in parallel with a tunneling path elsewhere in the film.

The literature [25, 26, 28] shows that the electric fields required for EFIE and EFISD are ~26 V nm$^{-1}$ [28] and ~1 V nm$^{-1}$ [25] respectively for tin. Since the electric field across a typical ~ 0.2 nm gap in our samples can exceed ~50 V nm$^{-1}$ at the peak ramped voltage (Fig. 3 (c)), it is clear that formation of an Sn wire (and consequent quantized conductance steps) is indeed possible by both EFIE and EFISD.

Having explained the *increases* in conductance in response to an applied field by formation of atomic scale wires, it is clear that *decreases* in conductance can be explained by breaking the same wires due to electromigration i.e. electric fields and high currents drive atomic motion, reducing the size of the conductor and eventually breaking it. [29] The current densities in our experiments can easily exceed those required for electromigration in tin ($2 \times 10^8$ A m$^{-2}$). [30] It is not surprising that electromigration causes a monotonic decrease in conduction in small samples where a single atomic scale wire dominates.

To confirm the effects described qualitatively above we have simulated conduction through a 2D percolating system. The system size for the data presented here is 200 x 200 particles and the particles are allowed to have random positions. This continuum model[9] is essentially the same as that of "2D penetrable disks"[20] except that we explicitly allow tunneling. This continuum percolating-tunneling regime has not previously been quantitatively modeled.

We focus here on the regime below the percolation threshold (see Fig. 1 (a)) where the conductance of the system is due to tunneling across gaps between the groups of connected particles, and connections between particles have negligible resistance. Each gap is assigned a conductance

$$G_i = A \exp(-\beta L_i) \tag{1}$$

where $A$ and $\beta$ are constants and $L_i$ is the size of the gap. The conductance of the network is then calculated by solving for the voltages at each node. [31] This model is valid so long as, in the experiments, the series resistance of the well-connected particles is small compared to $1/G_0$ ~12.9 kΩ. The resistances of a 40 nm Sn particle and a uniform Sn film with thickness ~20 nm are both estimated to be ≤ 20 Ω (the bulk resistivity of Sn is ~$10^{-7}$ Ω m$^{-1}$ [21]), and so this is indeed the case.

We simulate the switching process by identifying the gaps with the largest electric fields and replacing them with a quantum conductor ($G_i \rightarrow G_0$), thus simulating the formation of an atomic wire by EFISD or EFIE (Fig. 1 (e) – (f)). We then re-calculate the conductance of the network. Fig. 4 (a) shows a histogram of the system conductances before (blue bars) and after (red bars) creation of the quantum conductor for several hundred realizations of the system with $p$ = 0.675 (close to, but below, $p_c$). It is evident that a significant number of the trials result in switching to a total conductance close to $G_0$. Fig. 4 (b) shows that the fraction of trials resulting in a system conductance close to $G_0$ increases monotonically as $p \rightarrow p_c$. Therefore the simulations strongly support our model of formation and breaking of atomic-scale wires in response to an applied field.

We have also simulated the effect on sample conductance when atomic wires are formed during the onset of conduction (when only small electric fields are present). We identify the gaps with the highest tunneling currents as most likely to cause a jump to contact via van de Waals forces, and replace them with quantum conductances. We obtain data very similar to that in Fig. 4 i.e. near the percolation threshold we obtain quantized conductances for a large proportion of the samples, consistent with the experimental data in Fig. 1.

The above picture of formation and breaking of atomic scale wires is strongly supported by experimental data obtained after repeated cycling of the voltage ($V_{max}$ ~2-4V) when electromigration has broken all conductive connections in the film and the sample is in a low conductance state (similar to that for $p < p_c$, prior to the onset of conductance). We find that by then applying higher voltages (up to 10V) steps towards higher conductance are observed, as in Fig. 3 (c) i.e. switching behavior can be restarted. Here the high voltage induces switching to near quantized levels ($G_{max} \sim G_0/2$). This behavior is consistent with establishment of an atomic wire due to EFIE, and rapid subsequent breaking of the new connection due to the high currents that flow at these high voltages.

Fig. 3 (d) shows conductance data from Fig. 3 (c) on a log scale. The exponential increase in conductance at moderate voltages (>2V) is clear evidence that tunnel gaps dominate transport; temperature dependent data (not shown) is also consistent with tunneling. Furthermore, the switching behavior observed at room temperature (Fig. 3 (a)) ceases as the temperature decreases below ~200 K. This is consistent with an activated process such as EFIE / EFISD. These observations are entirely consistent with the model of tunneling and switching described above.

The behavior of these samples is complex (with switching between multiple states), as well as time dependent and history dependent. As such their stability is not sufficient for device applications at present. However, four out of 38 100x200 μm samples exhibited switching between well-defined on / off states for periods up to hours (Fig. 3 (b)). Active feedback controlled voltage supplies have proven helpful in controlled formation of nanogaps [29] and we expect that a similar approach could significantly improve reproducibility of the switching behavior.

Our sample characteristics are similar to those of memristors [8] and the inherent complexity suggests that neuromorphic behaviour [5,6] may be possible. Furthermore we believe that further interesting surprises in the properties of nanoscale percolating - tunneling devices should be expected. Certainly, a basic understanding of these devices is still lacking: even the question of whether or not there is a well-defined threshold for the onset of tunneling conduction remains unresolved [32]. Furthermore these devices provide an interesting route to furthering our understanding of novel insulating states formed from nanoscale superconductors, [33,34] and to fabrication of nanoscale gaps between pairs of electrical contacts that could be convenient for molecular electronics applications. [29]

**Acknowledgements**

The authors gratefully acknowledge financial support from the Marsden Fund, NZ, and the MacDiarmid Institute for Advanced Materials and Nanotechnology.

# Figures and Captions

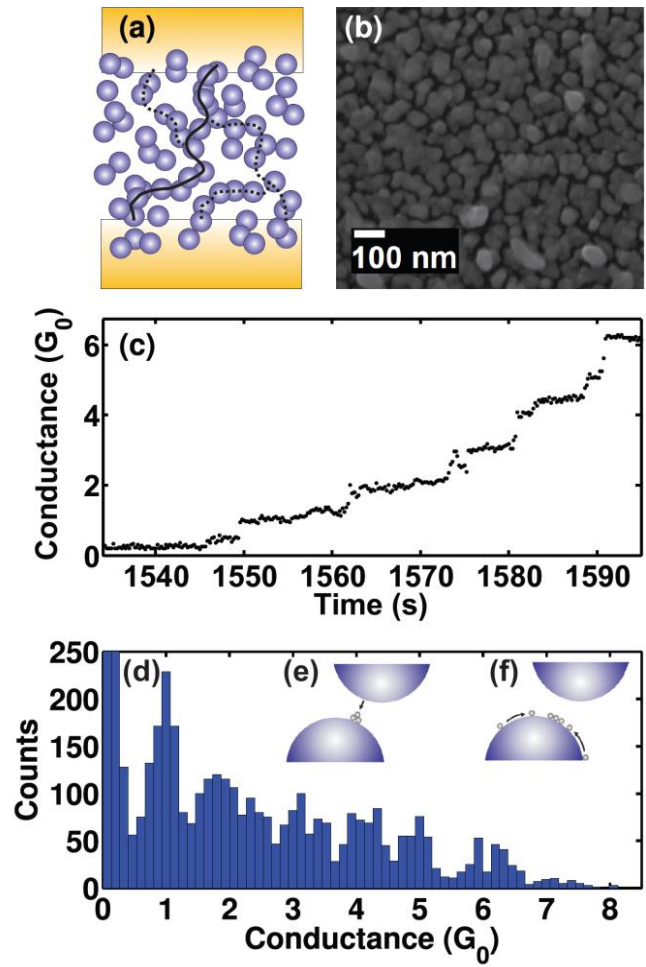

Fig. 1. (colour online) (a) Schematic illustration of a percolating tunnelling system showing a gap in the critical path (solid line), and additional tunnelling paths (dashed lines). (b) SEM micrograph and (c) onset of conductance as a function of deposition time for a typical sample. The steps in conductance coincide with integral multiples of $G_0$. (d) Histogram of conductance values measured during the onset of conductance in 21 samples, showing clear peaks at integral multiples of $G_0$. (e) Schematic depiction of electric field induced evaporation (EFIE) and (f) electric field induced surface diffusion (EFISD) or van der Waals forces.

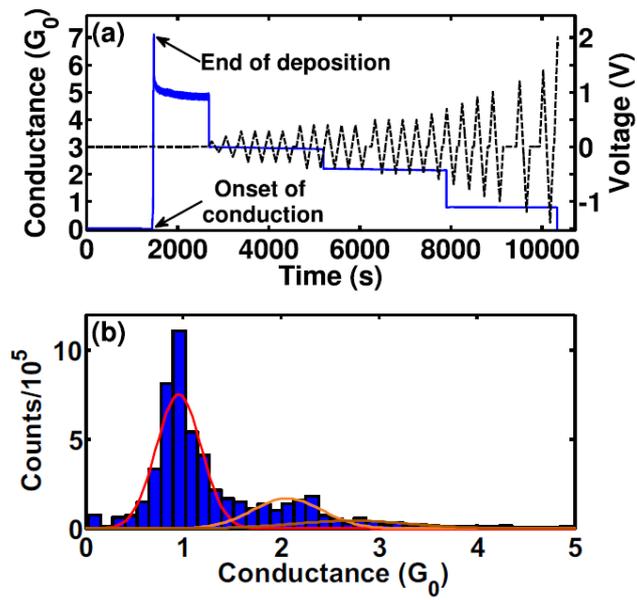

Fig. 2. (colour online) (a) Conductance of a 10 μm x 20 μm sample (solid line) showing downward steps under an applied voltage (dashed line). (b) Histogram of conductance data from all 67 samples showing a clear peak at $G_0$, and much weaker peaks near 2 $G_0$ and 3 $G_0$. All data was collected while ramping the applied voltage.

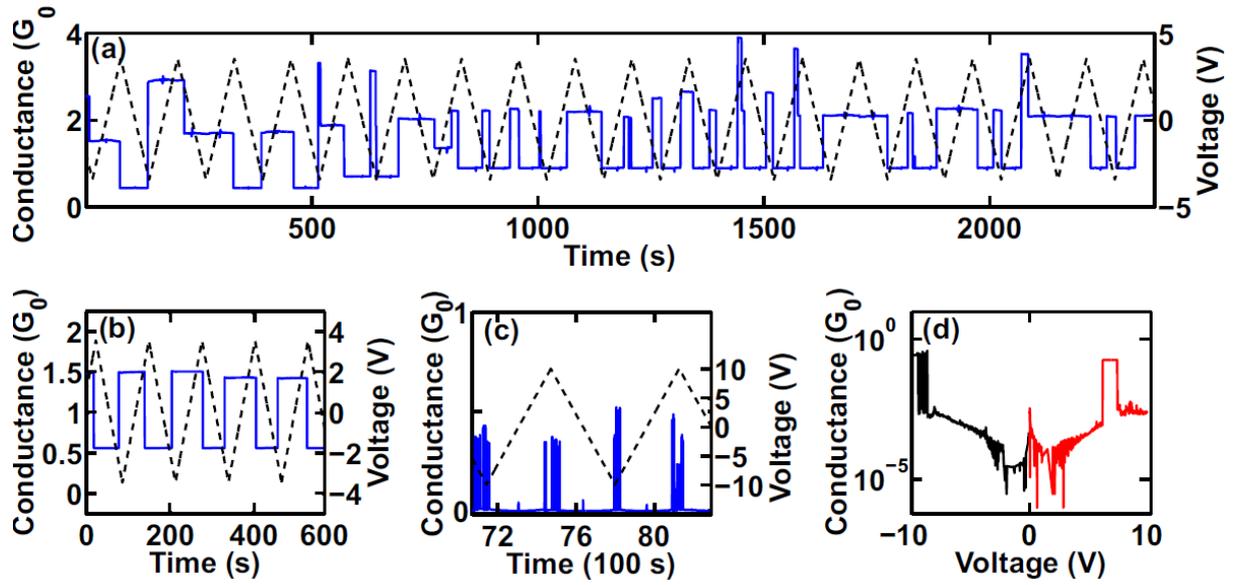

Fig. 3. (colour online) (a) Typical multi-level switching behavior under an applied voltage ramp. (b) Example of repeatable switching between the same two conductance levels. (c) Conductance data for a sample in the tunneling regime. (d) *I(V)* behavior for the same sample as in (c). The exponential increase is clear evidence for tunnel gaps. The data for $V \leq 2V$ is noise and should be ignored.

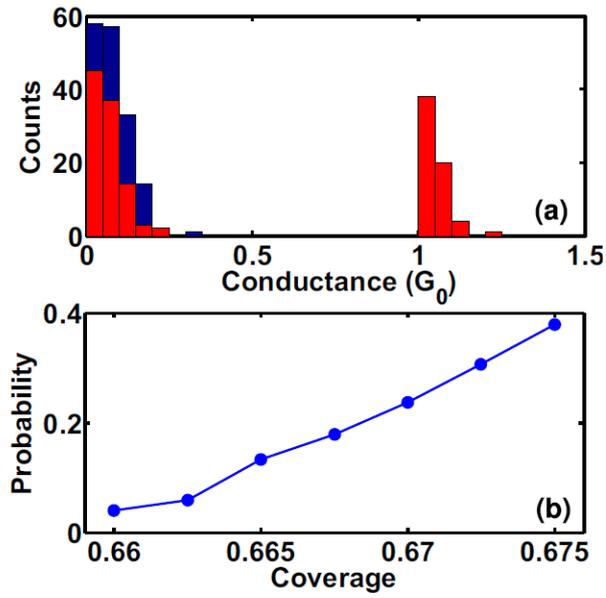

Fig. 4. (colour online) Results of continuum percolation model, for system size 200 x 200 particles and $A=1$ and $\beta=100$ (corresponding to a tunneling length scale of 0.01 particle diameters). (a) Histogram of *system* conductance values prior to (blue, darker) and after (red, lighter) insertion of a quantum conductance ($G=G_0$) in the gap with the highest electric field, for $p=0.675$. (b) Probability of observation of quantized conductance for the *system* as a function of $p$.